\documentclass[sigconf]{acmart}
\usepackage{listings}
\usepackage{listings}
\usepackage{booktabs}

\usepackage{array}
\usepackage{multirow}
\usepackage{enumitem}
\usepackage[linesnumbered,ruled,vlined]{algorithm2e}
\usepackage{appendix}
\setcopyright{none}
\renewcommand\footnotetextcopyrightpermission[1]{} 

\AtBeginDocument{%
  \providecommand\BibTeX{{%
    \normalfont B\kern-0.5em{\scshape i\kern-0.25em b}\kern-0.8em\TeX}}}

\graphicspath{{./images/}} 
\begin{document}
\pagestyle{plain}
\title{Template-Based Schema Matching of Multi-Layout Tenancy Schedules:A Comparative Study of a Template-Based Hybrid Matcher and the ALITE Full Disjunction Model}

\author{
  Tim Uilkema$^{1}$, 
  Yao Ma$^{1}$, 
  Seyed Sahand Mohammadi Ziabari$^{1,2}$,
  Joep van Vliet$^{3}$
}

\affiliation{%
  \institution{$^{1}$Informatics Institute, University of Amsterdam}
  \streetaddress{1098XH Science Park}
  \city{Amsterdam}
  \country{The Netherlands}
}

\affiliation{%
  \institution{$^{2}$Department of Computer Science and Technology, SUNY Empire State University}
  \city{Saratoga Springs, NY}
  \country{USA}
}

\affiliation{%
  \institution{
  $^{3}$CBRE, 1059 CM Amsterdam, The Netherlands}
  \country{}
}

\email{tim.uilkema@student.uva.nl, y.ma3@uva.nl, sahand.ziabari@sunyempire.edu, joep.vanvliet@cbre.com}


\begin{abstract}
The lack of standardized tabular formats for tenancy schedules across real estate firms creates significant inefficiencies in data integration. Existing automated integration methods, such as Full Disjunction (FD)-based models like ALITE, prioritize completeness but result in schema bloat, sparse attributes and limited business usability. We propose a novel hybrid, template-based schema matcher that aligns multi-layout tenancy schedules to a predefined target schema. 

\indent The matcher combines schema- (Jaccard, Levenshtein) and instance-based metrics (data types,  distributions) with globally optimal assignments determined via the Hungarian Algorithm. Evaluation against a manually labeled ground truth demonstrates substantial improvements, with grid search optimization yielding a peak F1-score of 0.881 and an overall null percentage of 45.7\%. On a separate ground truth of 20 semantically similar column sets, ALITE achieves an F1-score of 0.712 and 75.6\% nulls. These results suggest that combining structured business knowledge with hybrid matching can yield more usable and business-aligned schema mappings. The approach assumes cleanly extracted tabular input, future work could explore extending the matcher to support complex, composite tables. 
\end{abstract}
\maketitle
\section*{Github Repository}
\url{https://github.com/TUilkema}

\section{Introduction}
\label{sec:introduction}

Organizations increasingly rely on data-driven systems to support their decision-making. In the real estate industry, a significant amount of valuable information is stored in unstructured data formats such as PDFs, which often contain critical data for operations. An example is tenancy schedules, which are mostly stored in PDFs and formatted in tables. These documents use different table templates to store the same attribute information (Price, total area rented, start date lease). These documents are crucial for property management, financial reporting, and general business intelligence applications. However, manually extracting large quantities of data is impractical and highly time-consuming.

\indent Capturing relationships among heterogeneous datasets, commonly referred to as schema matching, is a fundamental challenge in data integration \cite{koutras_valentine_2021}. Schema matching methods fall into three main categories \cite{koutras_valentine_2021}:
\begin{itemize}
    \item \textbf{Schema-based approaches} which rely on metadata such as column names and data types (Cupid \cite{madhavan_generic_nodate}, COMA \cite{do_coma_2002}).
    \item \textbf{Instance-based approach} which analyze data values to determine similarity.
    \item \textbf{Hybrid approaches} which combine both schema and instance-based approaches.
\end{itemize}

Existing work has mostly focused on structured databases, with significant work in holistic or automated schema integration techniques.
Effectively processing unstructured data remains a key challenge in data-intensive industries, particularly due to the complexity and inconsistency of document formats \cite{mishra_structured_2017}.
A major obstacle in this domain is the diversity of layouts in unstructured documents, which complicates automated information extraction. 

\indent Several approaches have been developed to address schema matching and table integration, these focus on structured datasets. One recent framework, ALITE uses an approach that automatically merges tables from data lakes into one large unified schema while retaining all columns. \cite{khatiwada_integrating_2022}. Full Disjunction (FD) used in ALITE retains all columns and unifies tables into a large table. Unique and sparse attributes which are retained introduce schema bloat \cite{paulraj_ext-nosql_2024}. This FD method retains all information, however, it also introduces possible schema bloat with many columns containing null values. While this ensures no loss of information, it is not optimized for business applications, where a more concise and meaningful representation of critical attributes is required.

\indent A potential alternative is a template-based approach, where users define a target schema with all the essential columns. Traditional schema matching methods are used to map tables accordingly.
The template-based approach, introduced in this paper, explicitly allows for the definition of a extendable target schema, ensuring that only essential attributes are integrated, without requiring separate templates for each table layout. Current research highlights the difficulty of handling varied table structures and points out a gap in how existing methods struggle to generalize across different layouts and integrate extracted data into a unified structured format \cite{baviskar_efficient_2021}. 

\indent To address this gap, this research aims to introduce a template-based approach using schema matching methods to automate table integration. 

\indent This research focuses on comparing the FD baseline of ALITE with a novel template-based based hybrid matcher. This research aims to apply and adapt hybrid schema matching methods to unify tables. This research hopes to address the challenge of harmonizing diverse table layouts into a single structured format. The study will focus on aligning schemas into a unified output, with an emphasis on handling the diverse table layouts of different tenancy schedules found in the real estate business. 
A Data Requirements Document (DRD) specifying attributes and layout used in internally by CBRE is chosen as the target schema. The hybrid matcher combines schema-based similarity metrics (such as name and string similarity) and instance-based similarity metrics (such as value type and distribution similarity) to improve alignment accuracy. By integrating multiple matching metrics, the hybrid approach hopes to provide a more robust accuracy across varied table layouts. In doing so, this research addresses the challenge of automated matching in tenancy schedules of varying table layouts by using hybrid matching signals and a novel target schema.

This study investigates the effectiveness of a hybrid template-based schema matcher in aligning multi-layout tenancy schedules to a structured target schema, the DRD format. The research specifically compares this hybrid approach against a FD baseline, which retains all attributes across input tables at the cost of introducing redundancy and null-heavy schemas.

The proposed matcher combines both schema-based and instance-based similarity metrics and leverages target-specific metric weights to guide the matching process. The evaluation assesses not only the alignment accuracy—using F1 scores against a manually annotated gold standard—but also examines how grid search optimization of these metric weights influences matching performance. In addition, schema usability is evaluated through compactness and null density, offering insight into the practicality of the resulting integrated schemas in downstream applications.

The remainder of this paper is structured as follows. Section 2 surveys prior work on schema matching and table integration, highlighting key limitations of current approaches. Section 3 introduces the hybrid matcher and FD baseline, detailing their underlying mechanisms. Section 4 presents the evaluation methodology and experimental setup, followed by the results and discussion in Section 5. Finally, Section 6 concludes the paper with reflections on contributions and outlines directions for future research.

\section{Related Work}
\label{sec:related_work}
Recent research has focused on integrating heterogeneous tabular data from diverse sources, particularly in the context of data lakes and (semi or un)-structured documents. A recent survey emphasizes that different tasks such as table extraction, schema matching and schema augmentation are distinct yet related components of the tabular data augmentation pipeline. Each of these tasks has unique technical challenges and goals \cite{cui_tabular_2024}.
Approaches such as ALITE \cite{khatiwada_integrating_2022} and DeepJoin \cite{dong_deepjoin_2023} represent state-of-the-art methods for combining tables through schema matching. ALITE applies Full Disjunction (FD) to preserve all attributes across all documents, this ensures completeness but at the cost of producing a highly sparse schema populated with null values and redundant fields. In contrast, template-based schema mapping enables domain experts to define a standardized target schema that prioritizes all relevant business attributes. These methods typically require one template per document layout which highly affects the business applicability for diverse layout documents \cite{baviskar_efficient_2021}.An example of this is discussed in automating invoice extraction methods \cite{leon_extracting_2021}. Such methods may overlook semantically relevant but structurally inconsistent columns across layouts. 

\indent This review of the literature delves into the evolution of schema matching from manual and rule-based methods to learning-based approaches. This review identifies a gap in hybrid schema integration strategies that can reconcile the scalability of automated matching with the precision of template-based design. Specifically, we focus on the underexplored domain of tenancy schedule integration, where many distinct PDF tables represent similar data. We investigate how schema matching techniques can be optimized to balance completeness and minimality in multi-layout real estate documents.

\subsection{Table Extraction}
Table extraction is the process of identifying, interpreting, and structuring tabular data from documents into machine-readable formats such as CSV or relational data \cite{burdick_table_2020}. This task of extraction precedes other tasks such as schema matching and table augmentation \cite{cui_tabular_2024}. Table extraction can typically be divided into three sub-tasks, table detection, structure recognition and functional analysis \cite{chi_complicated_2019,smock_pubtables-1m_2022}.

\indent Early table extraction methods were rule-based, they relied on heuristics such as line detection, whitespace analysis or layout templates. These approaches worked well for regular table formats however, they struggled with diverse or noisy layouts and required manual steps per document type. Enterprise data is estimated to comprise of up to 80\% unstructured data out of all business data \cite{mishra_structured_2017}, this led to researchers developing more generalizable techniques. This includes machine learning approaches, such as conditional random fields (CRFs) and other statistical methods that infer table regions based on cues from the text or visual cues \cite{pinto_table_nodate}. Hybrid models combined layout features and statistical methods. Systems like OCR++ \cite{singh_ocr_nodate} have shown superior performance in extracting structured table elements from PDF documents. This highlights the difficulty of relying solely on rule-based or more layout-dependent approaches.

\indent The field has since shifted toward deep learning. Convolutional neural networks (CNNs) were first used to detect table boundaries in document images \cite{gilani_table_2017}. Later, models like TableNet \cite{paliwal_tablenet_2020} and GraphTSR \cite{chi_complicated_2019} proposed end-to-end pipelines for detecting and segmenting tables including complex spanning cells and nested headers.

\indent More recently we have transformer-based models which have become state-of-the-art. The release of PubTables-1M \cite{smock_pubtables-1m_2022}, a dataset of around one million annotated scientific tables, enables higher performance models by training on a large diverse dataset. These models include TableFormer \cite{nassar_tableformer_2022} and LayoutLMv3 \cite{huang_layoutlmv3_2022} and use self-attention mechanisms to capture textual and spatial structure information. This makes for robust models which can adapt to format variation and are suitable for multi-layout documents like tenancy schedules. Despite such recent advances, a single general-purpose table extractor can still fail on complex or specific domain layouts. This challenge is also found in invoice extraction where layout diversity demands specialized approaches beyond more general-purpose extraction methods \cite{saout_overview_2024}.

\indent This review highlights significant advances in table extraction technology while highlighting domain-specific challenges of extracting structurally diverse tables and how this is an underexplored area in the academic literature. The next field of research lies in schema matching; this is the act of linking extracted table content to one another or a unified schema.

\subsection{Schema Matching and Table Integration}
Schema matching is the process of identifying correspondences between attributes in different data schemas \cite{madhavan_generic_nodate}. This is a core task in data integration and is especially relevant when merging datasets with varying structures into a unified representation. In this research, we explore this field with regard to tenancy schedule integration, where real estate firms often use different schemas to represent similar lease-related data. Schema matching enables the alignment of columns/attributes which are semantically similar but lexically varied like ‘Tenant’, ‘Lessee’ or ‘Occupant’ to a common target attribute. Instance-based and embedding-based techniques support such alignments when names differ significantly \cite{nargesian_table_2018}. Over the past two decades, the field has evolved from simple string-based heuristics to composite matchers and machine learning approaches. However, existing methods still fall short when it comes to scalable, multi-layout document integration into a unified domain-specific schema.

\subsubsection*{Earlier Schema Matching Techniques}
\leavevmode\\
Early schema matching systems relied on heuristic, rule-based or structural techniques. One foundational system, Cupid \cite{madhavan_generic_nodate}, matched schema elements using linguistic similarity and structural context (parent-child relationship). Similarly, Learning Source Description (LSD) \cite{doan_reconciling_2001} introduced a learning-based matcher that combined multiple classifiers using features such as attribute names, data types and sample values.

\indent Another big advancement was made with Similarity Flooding (SF) \cite{melnik_similarity_2002}, this introduced a graph-based matcher approach. This system pioneered the difference between schema-based matchers (using only metadata) and instance-based matchers (using the data values). These  were effective for small-scale scenarios however, early systems did not scale well to many-to-many or many-to-one integration, which is typical in modern enterprise pipelines \cite{melnik_similarity_2002}.
\subsubsection*{Schema Matching for Table Integration at Scale}
\leavevmode\\
As schema matching advanced, large-scale integration frameworks started emerging. A notable system is COMA \cite{do_coma_2002} which allows users to configure and combine different similarity measures to provide a mixed matcher. The successor was COMA++ \cite{aumueller_schema_2005} and added support for reusing mappings. These systems support hybrid matchers which combine name similarity, structure and instance information.

\indent With the rise of big data and data lakes, systems began targeting integration at scale. Valentine \cite{koutras_valentine_2021} for instance offered a benchmarking framework to compare different matchers like COMA, Cupid and others across datasets. It showed that no single method can outperform all others in all contexts and that the dataset heavily influenced the accuracy \cite{koutras_valentine_2021-1}.

\indent Learning-based methods emerged as well with DeepJoin \cite{dong_deepjoin_2023} using pre-trained language models such as BERT to learn semantic representations of columns and use this to predict the joinability of columns. This was not designed for full schema integration, it did however demonstrate that deep models can align and join semantically related columns when names differ significantly. This is an important aspect of tenancy schedule integration as columns with the same attribute can differ significantly by name. Recent work demonstrates the use of LLMs for extracting structured entity-relation pairs from scientific texts in with domain-specific relationships. This is a powerful development in extracting structured data but does not deal with tabular data \cite{dagdelen_structured_2024}. 
Full-schema integration was achieved by ALITE using Full Disjunction (FD) \cite{khatiwada_integrating_2022}. This system retained all attributes from all input tables to avoid data loss. However, a downside of this approach is that it leads to a null-heavy schema with many redundant attributes and is impractical in real-world applications.

\noindent\text{Templates vs Automated Table Integration}.
In practical enterprise settings, schema integration often relies on template-based matching where a predefined target schema serves as the ground truth. Each incoming document layout is manually mapped to this schema \cite{khatiwada_integrating_2022} which tries to ensure consistency and introduces issues with scalability. This manual approach remains common in industry but becomes increasingly inefficient as document layout diversity grows \cite{madonna_university_unstructured_2016}. Recent work performed focuses on fully automated table augmentation pipelines from data lakes, this approach requires extensive preprocessing and join discovery \cite{cappuzzo_retrieve_2025}. Their result showed that even sophisticated automated merging leads to noisy and null-heavy tables when joins are imprecise. This supports the need for semi-automated mapping strategies using a template-based approach.

\indent Another recent study underscores the increasing inefficiency of manual schema mapping at a larger scale \cite{soperla_automated_2025}. This study underscores the inefficiency of manual schema mapping at scale and reveals that organizations spend over 70\% of data integration time on mapping and validating tasks with error rates of up to 9\%. The required maintenance tasks consumed tens of thousands of person-hours annually. These findings confirm that manual schema mapping per layout basis is not sustainable in large-scale environments. Manual mapping is still widely used and introduces bottlenecks when integrating schemas at scale \cite{asif-ur-rahman_semi-automated_2023}.

\indent Such systems typically require one template per layout which leads to maintenance overhead and poor scalability in multi-layout domains such as rent rolls. Academic systems like ALITE or Valentine do not consider this use case. Most assume that a fully automated integration is required (union all schemas and retain all attributes) or perform pairwise matching with individual fixed target templates.

\indent No current literature was found that robustly supported multi-layout-to-one schema integration, especially in the case of varying layouts and the requirement of a scalable semi-automated mapping system. Earlier research had the goal of fully automated integration and this is an underexplored use case \cite{rahm_survey_2001}.

\begin{table}[H]
\small
\renewcommand{\arraystretch}{1.1}
\small
\centering
\begin{tabular}{|c|p{2.2cm}|p{4.6cm}|}
\hline
\textbf{Year} & \textbf{Milestone} & \textbf{Description}\\
\hline
2001 & LSD & Early ML-based schema matching.  \\
\hline
2002 & Cupid / COMA / Similarity Flooding & Linguistic and structural matchers (Cupid), graph-based matching (SF), and composite matchers (COMA). \\
\hline
2005 & COMA++ & Reuse of prior mappings \\
\hline
2017 & TableNet& CNN-based table detection  \\
\hline
2018 & Table Union Search & Matching union-able tables at scale \\
\hline
2019 & GraphTSR& Complex table structure recognition \\
\hline
2020 & Valentine & Benchmarking framework for schema matching algorithms  \\
\hline
2021 & PubTables-1M  & One million annotated tables; boosted deep table extraction models \\
\hline
2022 & ALITE  & Full Disjunction for schema integration  \\
\hline
2023 & DeepJoin  & BERT-based semantic join-ability prediction \\
\hline
2024 & Tabular Augmentation & Use of generative AI for augmenting tabular training data \\
\hline
2025 & Retrieve, Merge, Predict & Automated table enrichment from data lakes  \\
\hline
2025 & Soperla & Empirical study: highlights manual bottlenecks\\
\hline
2025 & \textbf{This Research} & Template-based schema matching for multi-layout PDFs (tenany schedules)  \\
\hline
\end{tabular}
\caption{Timeline of related work}
\label{tab:timeline}
\end{table}

\subsection{Research Gap}
Despite extensive work on schema matching and integration of data, existing literature reveals a gap in scalable schema alignment for multi-layout documents such as tenancy schedules.

\indent Enterprise data pipelines often rely on template-based data integration where a predefined target schema is manually mapped to each incoming document layout \cite{khatiwada_integrating_2022}. Invoice systems also still rely on user-defined templates per document layout and face the inefficiency of the one-template-per-layout approach \cite{saout_overview_2024}. This approach, requiring a template per layout, becomes increasingly labour-intensive as the number of document variations grows. Manual integration is known to consume over 70\% of data processing effort and contributes to operational inefficiencies \cite{soperla_automated_2025}.

\indent Recently fully-automated integration systems were proposed such as ALITE, which prioritizes completeness (preserving all attributes from all layouts by FD). This ensures no data is lost and consequently results in null-heavy schemas which are not usable in our business setting. Literature confirms that the ALITE framework suffers from sparse attributes and requires substantial post-processing to extract the required information and provide actionable insights \cite{cappuzzo_retrieve_2025}.

\indent Notably, no prior work addresses the scenario of matching diverse table layouts into a single predefined schema in a robust manner. This provides a particularly useful use case in domains with high structural variability but strict schema requirements such as the real estate domain. Existing schema matchers either operate on pairwise mappings or aim for a generalized fully automated union rather than enabling one-to-many layout normalization into a unified predefined schema.

\indent This research aims to fill this gap by proposing a hybrid schema matching pipeline and comparing this to fully automated solutions. This approach combines the interpretability and business alignment of template-based systems with the generalization of automated matching methods. By leveraging instance-based and embedding techniques for column/attribute alignment, the proposed solution supports multi-layout-to-one-schema integration, providing a more scalable alternative to both manual and fully automated solutions.

\section{Methodology}
\label{sec:methodology}

This research proposes a hybrid, template-based schema matching pipeline which is designed to align multi-layout tenancy schedules. The pipeline aims to maximize automation while maintaining high accuracy and business usability by leveraging a target schema. A template-driven method was selected to try to minimize schema bloat seen in FD-based methods. The pipeline consists of several stages, ranging from target schema definition and data preprocessing to hybrid matching and evaluation. Each stage in this pipeline is designed to support the multi-layout nature of tenancy schedules while handling schema alignment with a predefined and business-approved target schema. A high-level overview of the methodology pipeline is shown in Figure \ref{fig:methodology_pipeline}. This research follows an experimental design, where the proposed pipeline is evaluated against a manually labelled ground truth. The evaluation is focused on schema alignment accuracy and schema usability. Alignment accuracy is measured in F1-score and compactness in total number of columns and resulting null percentage.

\begin{figure}[ht]
  \includegraphics[width=\linewidth, trim=5 0 5 0, clip]{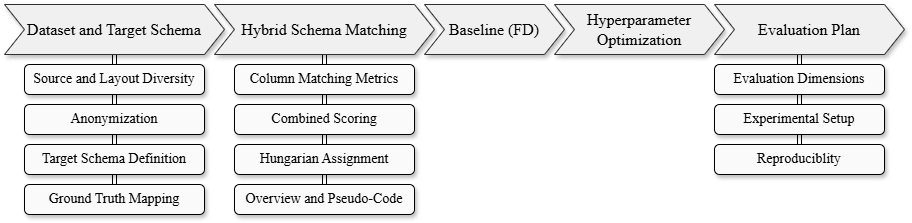}
  \caption{High-level overview of methodology}
  \label{fig:methodology_pipeline}

\end{figure}

Each of the following subsections provides insight into a specific stage of the pipeline, highlighting both the methodological choices made and the practical implementation details to support reproducibility.

\subsection{Dataset and Target Schema}
\subsubsection{Source and Layout Diversity}

To evaluate the robustness of the proposed hybrid schema matcher across real-world variability, a diverse dataset of tenancy schedules was compiled. This dataset includes documents from five distinct layouts: JLL, Savills, Park15, EDIF and CBRE. This dataset was sourced from the Capital Markets team at CBRE and represents actual documents exchanged between firms during property transactions. Tenancy schedules are highly confidential in nature and when exchanged fall under a NDA until the expiry date. For legal compliance, CBRE registers the NDA expiry dates associated with each document and propagates this to downstream business applications. This ensures that only permissible data is made accessible (for those limited documents which were manually extracted). The information stored in tenancy schedules is one of the most valuable sources of data for real estate advisory such as CBRE. CBRE provides advisory services in a data-driven manner. However, this data remains underutilized as the multi-layout nature prohibited data extraction until now. All tenancy schedules are of mainly office or multi-use business properties. The dataset accurately reflects multi-layout documents encountered in practice across office building tenancy schedules.

\indent Each layout captures tenancy data differently, some separate area usage into multiple columns of: 'Office Area' and 'Archive Area' while others provide a single 'Total Leased Area'. Similarly, some identical fields are expressed using different terminologies such as 'Lease Start Date' versus 'Commencement Date'. Differences in formatting are also present: monetary values, date formats, and surface area units (SQM versus $\text{m}^2$).

\indent This multi-layout nature exemplifies schematic heterogeneity where similar information is represented under different schemas. The concept was discussed early in a foundational study on heterogeneous data structures \cite{miller_using_1998}. Such heterogeneity poses a significant challenge for automated schema matching and motivates the need for a robust, layout-independent method. By including layout diversity throughout our dataset we measure the accuracy of aligning columns to a fixed target schema across varied source formats. By using this varied and representative set of tenancy schedules, this study tries to ensure that matching accuracy is evaluated not only on semantic similarity but also on the ability to generalize across real-world schematic variance.

\subsubsection{Anonymization}

Tenancy schedule data is inherently sensitive and contains confidential business information such as tenant identities, rental prices, and detailed lease terms. To ensure compliance with data privacy obligations and protect commercially sensitive information, all data used in this study was anonymized prior to experimentation.

\indent The anonymization process was carefully designed to preserve structural integrity, formatting-  and statistical characteristics of the original data while obscuring personally identifiable and commercially sensitive content. This ensures the anonymized dataset remains suitable for schema matching tasks while protecting confidential information. The following methods were applied to achieve this:

\begin{description}
    \item[Tenant Names] Real tenant names were replaced with randomly selected corporate names from external sources, such as company.info or synthetically generated using realistic naming patterns. These included combining typical corporate suffixes (B.V., N.V., GmbH, Ltd.) with selected business-like names.
    \item[Monetary Values] Rental prices and other financial information were randomized centred around a realistic office market mean rate of €250 per square meter with a standard deviation of €40 - € 50 per format. The original currency formatting was preserved for each layout (€ 201.235,88 versus  201235.88).
    \item[Leased Area] Surface area values were replaced with numbers randomly generated numbers within realistic bounds based on average leased areas per usage type (office, archive, storage).
    \item [Dates] All date fields were shifted consistently while preserving relative relationships (the time between the lease start date and lease end date remained the same). The formatting remained distinct per layout (2015-01-19, 19 jan 2015, 19 jan 15). In cases where dates consistently started on the first day of a month or year this structure is respected.
\end{description}
This anonymization approach allows us to perform real-world applicable schema matching while ensuring that no confidential data was used. We preserve original data types, numeric distributions and layout characteristics. All anonymized data was manually reviewed to ensure adherence to the original structure.

\subsubsection{Target Schema Definition}

A foundational element of the novelty of this template-based hybrid pipeline is the use of a predefined target schema instead of fully automated schema alignment. In contrast to other template-based data integration methods this proposed method does not require one template per layout but requires a target schema per unified schema layout.
This target schema acts as an anchor for aligning diverse tenancy schedule formats. This study adopts an existing, business-approved target schema from CBRE's internal Data Requirements Document (DRD). This DRD is used in practice to define how tenancy data should be recorded in Snowflake tables. The DRD schema defines a comprehensive set of attributes grouped across different entities; Property Identifiers (Building ID, Property Name, Address, City), Lease (Tenant Name, Unit, Use Type, Status), Financial (Annual Rent, Service, VAT), and Time (Commencement Date, Break Option, Expiry Date).

\indent By using this schema we ensure that aligned schemas comply with the information requirements of CBRE and avoid capturing attributes that are irrelevant or unused in downstream applications. This target schema serves two main purposes. First, it provides us with a fixed and standardized representation into which we must transform all tenancy schedule data. Attributes are defined in a clear manner such that downstream applications are not confused by the definition of attributes. Second, it allows the matching algorithm to work in a template-based manner where the system focuses on mapping source columns to a defined attribute rather than performing unsupervised column merging.

\indent This method directly addresses the challenge of scalability in data integration systems where new layouts require manually created extraction templates or one-off mappings. Instead, this pipeline relies on the fixed DRD schema and uses schema matching techniques to automatically align incoming documents regardless of layout. This reduces the onboarding time of new formats and enables extraction independent of layouts. Furthermore, this target schema can easily be extended with new attributes when required.

\indent The DRD target schema was converted into a machine-readable YAML format. Target attributes are enriched with signals such as synonyms, data types and typical statistical characteristics. For example the attribute ''commencement\_date'' may include synonyms like ''Start Date Lease'' or 'Lease Start'' and is defined as a date type with an associated date range. These typical date ranges can help differentiate the start and end date. Similarly, ''passing\_rent\_per\_pa'' includes synonyms such as ''Total Rent'' and ''Annual Rent'', it  is marked as a numerical type with typical ranges between '€50.000' and '€2.000.000'. These characteristics can help disambiguate cases where a source column like ``Office Total'' might appear similar to multiple target attributes such as ``Office Rent'' and ``Office Area''. We hypothesize that instance-based metrics can be a key discriminator by using value distributions. For instance. office areas typically range from 200 to 5,000 square meters, while rental values typically fall between €50,000 and €2.000.000. Statistical patterns can help resolve ambiguous mappings where schema-based methods fall short.

\indent The availability of instance-level characteristics, such as typical numerical ranges and distributional properties opens the door for fine-tuning the matcher through optimization. This study explores the impact of optimizing the relative weights of schema- and instance-based similarity components using a grid search strategy. The matcher can learn to emphasize different signals depending on attribute type, which evaluates whether optimized hybrid weighting improves matching performance as measured by F1-score.

\indent Importantly, this DRD schema has already been validated by domain experts and formally approved by CBRE's Central Europe Data Governance Lead. The adoption of this ensures that the final output of the pipeline is immediately usable by the business. It acts as a bridge between technical extraction processes and business reporting needs, enforcing both the reliability and usability of the resulting standardized data.

\subsubsection{Ground Truth Mapping}

To evaluate the effectiveness of the proposed schema matching pipeline and to support the potential calibration of matching strategies, a manually created ground truth was developed. This involved manually mapping each column of each layout form to the corresponding target attribute as defined in the target schema. All attributes are also correctly normalized into the desired formats. Essentially, for every source schema format we produce the correct alignment to the target schema.

The manual mapping process was performed using the attributes as defined in the DRD. For example, if a tenancy schedule format has columns labelled: 'Tenant Name', 'Level', 'Sqm Office', 'Sqm Archive', 'Start Date', these would be mapped to: tenant, floor, area office, area archive, commencement date. We make use of the DRD's definitions to help mitigate ambiguities and help identify attributes to deliberately omit. 

Ground truth mapping serves two purposes. First, it is the basis for supervised learning or rule calibration in our hybrid matcher. We can use the mappings created to train a model or fine-tune similarity thresholds. Secondly, the ground truth is essential for our evaluation stage. It allows us to quantitatively measure the correctness of the matcher on different metrics.

\subsection{Hybrid Schema Matching}
This section  describes the core algorithmic component of the pipeline: the hybrid schema matcher. Unlike more automated approaches such as Full Disjunction (FD), this matcher is template-driven. Our pipeline leverages both schema-based and instance-based signals to produce accurate column-to-attribute mappings. Each tenancy schedule is processed independently and matched to the target schema. The matcher follows a three-step sequence for each table: compute column-to-attribute similarity scores, combine scores, assign columns to attributes.

Figure \ref{fig:methodology_pipeline} provides a detailed illustration regarding the pipeline and column matching.

\subsubsection{Column Matching Metrics}

To support accurate mapping of source columns to target attributes, multiple similarity metrics are computed for each pair of column-to-attribute. These metrics are grouped into schema-based and instance-based categories.
\begin{itemize}
\item Schema-based
\begin{description}
    \item[Levenshtein Distance] Measures edit distance between column and attribute on a character level. This is useful in comparing similar strings such as: ``Commence Date'' and ``Commencement Date''
    \item[Jaccard Token Overlap] Measures set-based token similarity (``Tenant Name'' versus ``Name of Tenant'')
    \item[Synonym Matching] Synonyms defined in the target schema are used to enhance alignment in the schema-based metrics.
\end{description}
\item Instance-based
\begin{description}
    \item[Numeric Data Type] Calculates the likelihood that a column represents numerical values. Used to determine if a column such as ``Floor'' with values: ``3rd, 4th'' is a numerical column.
    \item[Date Type] Normalizes and verifies the values of the column to verify if it is of date values. 
    \item[Numerical Range Comparison] Compare numeric values against expected ranges.
    \item[KS-Test] Compares the distribution of the source column to the target attribute's distribution to evaluate similarity.
\end{description}
\textit{*Detailed explanation and formula for each metric are provided in Appendix \ref{app_metrics}}
\end{itemize}

\subsubsection{Combined Scoring}

The matcher computes a similarity score for each column-attribute pair using a weighted average of schema and instance metrics:
\[\text{Score}_{\text{total}} = \alpha \cdot \text{Score}_{\text{schema}} + (1 - \alpha) \cdot \text{Score}_{\text{instance}}\]
Here, \( \alpha \) is a tunable parameter to fine-tune the weights of schema versus instance scores. A column is only able to match to a potential attribute if the highest score exceeds the set threshold (threshold = $\theta$, default: 0.5). If a column does not have a sufficiently high similarity score it is explicitly omitted. Due to the multi-layout nature of tenancy schedules, some formats contain attributes which are highly specific and only appear in limited formats across the dataset. These rare attributes are too layout-specific to be meaningfully aligned across other formats. By mapping to a target schema, only relevant attributes (as determined by CBRE) are integrated, resulting in a business-relevant schema.

\subsubsection{Hungarian Assignment}

Once similarity scores are computed between all column-to-attribute pairs, the next step is to determine the best overall assignment. This is formulated as a maximum-weight bipartite assignment problem \cite{korula_maximum_2010}, where each column can be matched to at most one attribute.

\indent To solve this we apply the Hungarian Assignment Algorithm over the combined score matrix. Each cell $S_{ij}$ represents the total summed similarity score between source column $i$ and target attribute $j$ based on the combined scoring formula. The goal is to assign the best combination of columns to attributes that maximize the total sum of similarity scores across all pairs.

\indent A naive or greedy assignment approach would assign each column to its highest-scoring attribute independently. However, this often leads to suboptimal global assignment especially in the presence of ambiguous columns which were present in tenancy schedules. For example, multiple columns may have similarly high scores for a single target attribute. Greedy matching would assign the first one it encounters potentially preventing a better global alignment.
The Hungarian algorithm avoids this by considering the entire score matrix simultaneously and selecting the best combination of assignments to yield to highest total similarity score. A simplified example of the assignment problem is given in Table \ref{Hungarian Example}.

\begin{table}[ht]
\centering
\caption{Similarity score matrix for optimal matching using the Hungarian algorithm.}
\label{Hungarian Example}
\begin{tabular}{lccc}
\toprule
\textbf{Source} & \textbf{Office Area} & \textbf{Annual Rent} & \textbf{Commencement} \\
\midrule
Total & 0.95 & 0.94 & 0.12 \\
Rent & 0.93 & 0.60 & 0.17 \\
Start & 0.20 & 0.30 & 0.98 \\
\bottomrule
\end{tabular}
\end{table}

A greedy algorithm would result in a total score of $0.95+0.6+0.98 = 2.53$, whereas the globally optimal assignment achieves a total of $0.93+0.94 + 0.98 = 2.85$.

\subsubsection{Overview and Pseudo-Code}

The following Algorithm \ref{alg:matcher} outlines the full column-to-target-attribute matching process described above. For each extracted table column it is scored against every target attribute in the template schema. The matching is done using a combination of schema-based and instance-based metrics. The final assignment to target attributes is determined via the Hungarian Matching Algorithm with a threshold.

\begin{algorithm}
\caption{Column-to-Target-Schema}
\label{alg:matcher}
\KwIn{DataFrame $T$ from CSV with headers $H$; target schema attributes $S$; metric weights $w_m$; alpha $\alpha$; threshold $\theta$}
\KwOut{Mapping $M: H \rightarrow S$}
\ForEach{header $h \in H$}{
    \ForEach{attribute $a \in S$}{
        $s_{\text{schema}}   \leftarrow \sum_{m \in M_{\text{schema}}} w_m \cdot m(h, a)$ \;
        $s_{\text{instance}} \leftarrow \sum_{m \in M_{\text{instance}}} w_m \cdot m(T[h], a)$ \;
        $s_{\text{hybrid}}   \leftarrow \alpha \cdot s_{\text{schema}} + (1 - \alpha) \cdot s_{\text{instance}}$ \;
        \If{$s_{\text{hybrid}} \geq \theta$}{
            $C[h, a] \leftarrow s_{\text{hybrid}}$ \;
        }
        \Else{
            $C[h, a] \leftarrow -\infty$ \;
        }
    }
}

$(\mathit{rows}, \mathit{cols}) \leftarrow \text{Hungarian Algorithm}(-C)$ \;
$M \leftarrow \{ h_i \mapsto a_j \mid (i, j) \in (\mathit{rows}, \mathit{cols}) \ \wedge \ C[i, j] \neq -\infty \}$ \;
\Return $M$
\end{algorithm}
\ref{fig:methodology_pipeline}.
\begin{figure*}[htbp]
  \includegraphics[width=\linewidth]{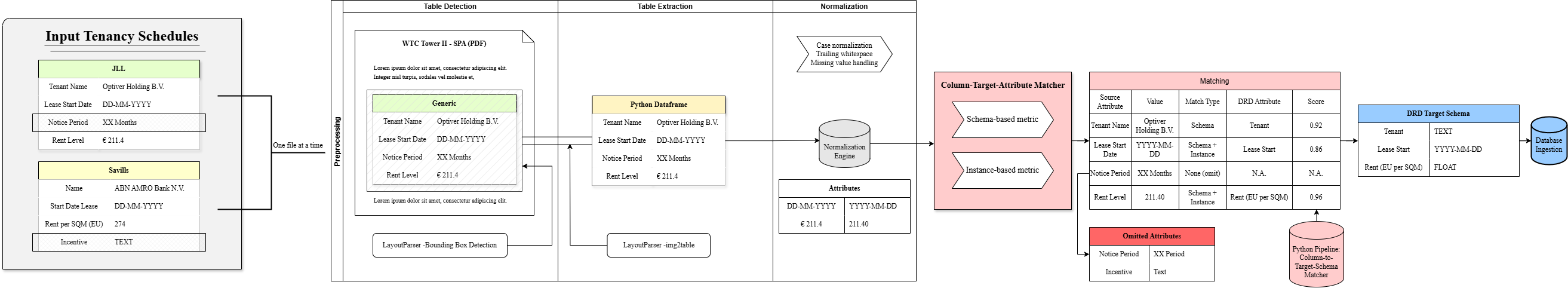}
  \caption{Detailed illustration of schema matching pipeline}
  \label{fig:methodology_pipeline}
\end{figure*}

\subsection{Baseline (FD)}
To assess the effectiveness of the hybrid matching approach, a baseline method inspired by the ALITE framework's use of Full Disjunction (FD) is utilized. Full Disjunction is an operation from database theory that informally combines all tables such that it retains all rows from any input, merging rows that share common information and filling nulls where information is missing. Conceptually, FD can be seen as a series of joins across all tables on all possible matching keys, ensuring that if any two records from different tables relate to the same entity, they end up in one combined record in the result. An FD model following the ALITE framework is used as a baseline as it has also been adopted in recent comparative studies for automated table integration\cite{ji_table_2025}. Using this method we theoretically merge all tenancy schedules into a single schedule which prioritizes data preservation with many redundant attributes, this can result in an extensive bloated schema.

\indent Essentially the baseline approach will outer-join all tenancy schedules, merging columns it identifies as join-able based on either exact string matches or semantic similarity. A BERT-based matcher is used to compute similarity by embedding column headers with values and attribute names. While the original ALITE framework uses a TURL-based model, this work adopts a BERT-based matcher for reasons of reproducibility and implementation simplicity.

\indent This research chooses FD as the baseline because ALITE's authors demonstrated that using FD as the integration mechanism can produce a more complete and unified dataset compared to simpler approaches, provided that schema matching is done correctly. By using FD without effective schema matching, we simulate a scenario of an integration done with poor alignment. This represents the scenario where all join-able formats in a data lake are integrated without domain knowledge and without considering a business-first approach.

\subsection{Hyperparameter Optimization}
The effectiveness of the hybrid schema matcher is influenced by several tunable hyperparameters. These include all the metric weights, the blend factor $\alpha$, which controls the relative importance between schema versus instance similarity, and the threshold $\theta$, which controls whether a similarity score is high enough to be able to assigned to a column-to-target pair (Hungarian Algorithm considers all pairs above threshold). These parameters are not fixed, they can be optimized to improve the overall matching performance.

\indent To systematically explore the optimal settings for these parameters, this research utilizes a grid search optimization using the ground truth mapping. The optimization can be done in a global configuration which finds the optimal settings for each parameter overall or on a attribute-specific level. This per-attribute optimization allows the matcher to adapt its behaviour to the nature of each attribute. This is based on the hypothesis that different attributes benefit from different matching signals. For example, numerical attributes such as ``Total Rent'' are more effectively matched using instance-based metrics, whereas textual attributes such as ``Tenant Name'' may rely more on schema-based metrics.

\indent The objective of this hyperparameter optimization is to maximize the F1-score across all tenancy schedules in the dataset. This subsection focuses on exploring the improvement in F1-score by conducting grid optimization on attribute-specific weights. By optimizing on a per-attribute basis, the matcher becomes more attuned to the multi-layout nature of real-world schemas and enables better generalization with limited ground truth mapping.

\indent However, performing a grid search is computationally expensive, particularly when evaluating all metric combinations for each attribute and document. To mitigate this, all metrics scores for each column per document are precomputed in a lookup tensor. This tensor allows the grid search to rapidly retrieve metric scores without recomputing them for each grid combination. This approach significantly reduced computational overhead and enabled a finer-grained grid interval.

\subsection{Evaluation Plan}
The evaluation of this research will focus on determining the feasibility and effectiveness of a novel template-based approach to schema matching. 

\indent The feasibility of the proposed approach is evaluated based on its implementation complexity and adaptability to a diverse range of tenancy schedule layouts. Effectiveness is assessed through a combination of accuracy, data completeness and usability, particularly in terms of producing a compact unified schema suitable for downstream applications.

\indent The evaluation of the proposed hybrid schema matcher focuses on the dimensions: of accuracy and usability. Accuracy is measured using standard classification metrics; precision, recall and F1-score. This is done by comparing results to a manually created ground-truth mapping. Usability is assessed by comparing schema compactness (amount of columns in the aligned schema) and overall null percentage.

\indent The ALITE baseline does not use a target schema and merges columns automatically using outer joins over column embeddings. To ensure fair usability comparison we created a secondary ground truth labeling of 20 semantically similar column sets. See Appendix \ref{tab:fd-mapping-example} for the semantic column groupings used in the baseline evaluation. This ground truth consists of sets ranging from 2 to 5 equivalent source columns. This allows us to compute F1-score, precision, and recall for ALITE. This is not done for direct comparison but to verify that its alignment accuracy is acceptable, thereby making usability comparisons meaningful. An F1-score above 0.7 was considered acceptable for the baseline to ensure that usability metrics such as null percentage reflect reasonably correct alignments. The schema usability of ALITE is then compared directly to that of the proposed hybrid matcher to address the evaluation of their performance.

\indent By analyzing these factors, this evaluation plan will assess whether a user-defined template-based approach provides a more efficient and business-aligned solution for rent roll schema matching than a fully automated integration method like ALITE.

\subsubsection{Experimental Setup}
\indent To assess the hybrid matcher we test four different configurations. The first is a default setup which uses equal weighting for schema- and instance-based metrics and default parameters of $\alpha = 0.5$, $\theta = 0.5$. The second configuration applies global parameter optimization where we tune $\alpha$ and $\theta$ (resulting in 0.75 and 0.51 respectively) based on the global F1 score. In this second setup, we keep all metrics at equal weights. Our third configuration introduces per-attribute weight optimization with the tuned parameters $\alpha$ and $\theta$ at 0.75 and 0.51. For each target attribute, we perform a grid search against the ground truth mapping where we optimize all weights in the range $[0, 1]$. In the final fourth configuration, we reset parameters to base values to isolate the effects of weight optimization alone and evaluate against parameter optimization.

\indent The FD baseline performs schema integration according to the ALITE framework and is tested in two different configurations using two transformed-based embedding models; 'all-mpnet-base-v2', and 'sentence-t5-base'. 

\subsubsection{Reproducibility}

All code, and evaluation scripts, ground truth mappings, target schema and source tenancy schedule formats are available at: 
\url{https://github.com/TUilkema}
For reference, partial target schema, ground-truth mappings and a sample of all source tenancy schedule formats including headers are found in appendix \ref{sec:apx:app_mapping}.

\section{Evaluation and Results}
This section presents the performance of the proposed hybrid schema matching pipeline, evaluated on our dataset and compared against two baseline configurations of ALITE. We first describe the dataset and manual ground-truth mappings used for benchmarking these results. Then, we assess model performance in terms of alignment accuracy and schema usability across several matcher configurations. We also examine how optimizing per-attribute metric weights differs between data types. Finally, we provide an error analysis to highlight common failures and  potential future improvements.
\begin{table*}[]
\centering
\caption{Comparison of hybrid matcher configurations vs ALITE baseline}
\label{tab:results}
\begin{tabular}{lccccc}
\toprule
\textbf{Method} & \textbf{F1} & \textbf{Precision} & \textbf{Recall} & \textbf{Compactness (\# columns)} & \textbf{Null \%} \\
\midrule
Hybrid (baseline: $\alpha=0.50,\ \theta=0.50$) & 0.778 & 0.789 & 0.767 & 17 & 47.8\% \\
Hybrid (optimized params: $\alpha=0.75,\ \theta=0.51$) & 0.800 & 0.795 & 0.806 & 17 & 50.1\% \\
Hybrid (optimized params + weights) & 0.875 & 0.913 & 0.840 & 17 & 45.9\% \\
Hybrid (optimized weights only, $\alpha=0.50,\ \theta=0.50$) & \textbf{0.881} & \textbf{0.926} & \textbf{0.840} & \textbf{17} & \textbf{45.7}\% \\
\midrule
\multicolumn{6}{l}{\textit{ALITE Variants}} \\
\quad ALITE (all-mpnet-base-v2)* & 0.699 & 1.000 & 0.538 & 89 & 75.9\% \\
\quad ALITE (sentence-t5-base)*\ & 0.712 & 1.000 & 0.553 & 88 & 75.6\% \\
\bottomrule
\textit{*Evaluated on manual ground truth of 20 column sets. See Appendix \ref{tab:fd-mapping-example}}
\end{tabular}
\end{table*}

\subsection{Dataset and Ground-Truth-Annotation}
This research uses a manually curated dataset of tenancy schedule PDF documents from five major real estate firms (4-6 tenancy schedules per format) to propose a pipeline for schema matching with heterogeneous table layouts. Each firm uses a distinct table layout, capturing the diversity of document formats in practice. The dataset has been fully anonymized while preserving data structure, syntax and formatting. Sensitive business data, such as tenant names and rent values have been changed to generic values. The five formats used here were: JLL, Savills, CBRE, Park15 and EDIF.

\indent Ground-truth mapping \ref{tab:fd-mapping-example} was created by manually aligning the tenancy schedules with the target schema (created with the help of the CBRE Data Requirements Documents) and grouping 20 semantically similar column sets manually. These serve as a benchmark for evaluating the accuracy of schema matching and training individual metric weights (instance- vs schema-based).

\subsection{Model Performance}

We evaluated the alignment accuracy of the template-based hybrid matcher using precision, recall, and F1-score, benchmarked against manually curated ground-truth mappings. In addition, schema usability was assessed based on schema compactness. The F1-score of the FD baseline is not directly compared in terms of alignment accuracy, but is included to ensure that alignment quality is sufficient for a valid schema usability comparison.

\indent We also compared different configurations of the matcher, including setups with and without per-attribute optimization. For each configuration, precision, recall, and F1-score were reported to quantify matching accuracy. Precision reflects the proportion of correct predicted matches, while recall indicates how many true column-to-target mappings are included in the standardized schedule. The F1-score, representing the harmonic mean between precision and recall, served as the main metric for optimizing parameters and weights.

\indent To assess schema usability from a business perspective, we compared two configurations of the FD baseline with those of the template-based matcher. Usability was measured using two indicators: the total number of columns in the standardized schema (schema compactness) and the overall null percentage. A compact schema with fewer nulls is considered more interpretable and usable for downstream applications.

\indent Table~\ref{tab:results} summarizes the performance of all four matcher configurations compared to the FD baseline. The baseline hybrid configuration achieved an F1-score of 0.778, while applying global parameter optimization increased this to 0.800. The highest performance was achieved with per-attribute weight optimization, reaching an F1-score of 0.881. This suggests that fine-tuning metric weights at the attribute level provides a significant improvement in alignment accuracy.

\indent Although we do not evaluate the alignment accuracy of the FD baseline directly, we use its F1-score to verify whether usability metrics can be meaningfully compared. The FD baseline reached an F1-score of 0.712 and a null percentage of 75.9\%, indicating more schema bloat compared to all hybrid configurations. This supports the hypothesis that the FD approach results in reduced schema usability.

\subsection{Attribute-Level Analysis} 
To further support the analysis of hyperparameter optimization, we analyzed the learned relative weights for schema- and instance-based metrics across the three attribute types of STRING, DECIMAL, and DATE.
Table \ref{tab:metric_gap} shows the average weight of schema- and instance-based metrics after per-attribute weight optimization using $\alpha=0.5,\theta=0.5$. 
\begin{table}[ht]
\centering
\caption{Average metric weight by data type}
\label{tab:metric_gap}
\begin{tabular}{lccc}
\toprule
\textbf{Type} & \textbf{Schema Avg} & \textbf{Instance Avg} & \textbf{Gap} \\
\midrule
STRING & 0.1500 & 0.2579 & 0.108 \\
DECIMAL & 0.1143 & 0.292 & 0.178 \\
DATE & 0.1000 & 0.356 & 0.256 \\
\bottomrule
\end{tabular}
\caption*{\textit{Weights are normalized to account for amount of metrics}}
\end{table}
The results indicate that instance-based metrics received constantly higher weights than schema-based metrics across all attribute types. The largest gain is observed for DATE attributes, where instance-based metrics outweighed schema metrics by a margin of 0.256. DECIMAL attributes also substantially benefit from instance-based features. This is likely because numerical distributions such as rent and square meters can have ambiguous header names where numerical signals are a key discriminator. This supports the idea for per-attribute weight tuning as it allows the hybrid matcher to adapt to different attribute characteristics and better use the informative signals. STRING attributes showed the smallest gap, likely due to due absence of numerical patterns to discriminate mappings. To evaluate the statistical difference between schema- and instance-based metrics, a Wilcoxon signed-rank test was performed on the metric weight configuration from the grid size 4 tuning set. The test yielded a test statistic of 3.0 with a p-value of 0.00049, this indicates a statistically significant preference for instance-based metrics.

\subsection{Error Analysis}
Despite the overall matching accuracy there were some recurring misalignments. One frequent example is the confusion between semantically similar attributes such as ``Office'', ``Office Area'', and ``Office Rent''. Without per-attribute tuned weights the matcher may incorrectly prioritize schema-based similarity (string synonym overlap) and map ``Office'' to ``Office Rent'' rather than the correct target of ``Office Area''. Per-attribute weights would reveal that the columns denotes ''Office Area'', as it fits better into the numeric distribution signals of this target attribute. This issue illustrates the risk of relying only on generic string-based metrics.

\indent These kinds of errors come to light even more in the Hungarian algorithm step. Here the algorithm seeks a globally optimal one-to-one mapping. However, even a small scoring imbalance (the difference between office\_rent and office\_area in the score is minimal) can lead to systematic mismatches when competing attributes have overlapping names. 

\indent In addition to matching errors, we observed that the overall null percentage can be a misleading statistic as there are attributes that are rarely present in any source tenancy schedules. These contribute to null values across all configurations tested. Future work could isolate new nulls introduced instead of overall null percentage.

\section{Evaluation and Results}
This section presents the performance of the proposed hybrid schema matching pipeline, evaluated on our dataset and compared against two baseline configurations of ALITE. We first describe the dataset and manual ground-truth mappings used for benchmarking these results. Then, we assess model performance in terms of alignment accuracy and schema usability across several matcher configurations. We also examine how optimizing per-attribute metric weights differs between data types. Finally, we provide an error analysis to highlight common failures and  potential future improvements.
\begin{table*}[]
\centering
\caption{Comparison of hybrid matcher configurations vs ALITE baseline}
\label{tab:results}
\begin{tabular}{lccccc}
\toprule
\textbf{Method} & \textbf{F1} & \textbf{Precision} & \textbf{Recall} & \textbf{Compactness (\# columns)} & \textbf{Null \%} \\
\midrule
Hybrid (baseline: $\alpha=0.50,\ \theta=0.50$) & 0.778 & 0.789 & 0.767 & 17 & 47.8\% \\
Hybrid (optimized params: $\alpha=0.75,\ \theta=0.51$) & 0.800 & 0.795 & 0.806 & 17 & 50.1\% \\
Hybrid (optimized params + weights) & 0.875 & 0.913 & 0.840 & 17 & 45.9\% \\
Hybrid (optimized weights only, $\alpha=0.50,\ \theta=0.50$) & \textbf{0.881} & \textbf{0.926} & \textbf{0.840} & \textbf{17} & \textbf{45.7}\% \\
\midrule
\multicolumn{6}{l}{\textit{ALITE Variants}} \\
\quad ALITE (all-mpnet-base-v2)* & 0.699 & 1.000 & 0.538 & 89 & 75.9\% \\
\quad ALITE (sentence-t5-base)*\ & 0.712 & 1.000 & 0.553 & 88 & 75.6\% \\
\bottomrule
\textit{*Evaluated on manual ground truth of 20 column sets. See Appendix \ref{tab:fd-mapping-example}}
\end{tabular}
\end{table*}

\subsection{Dataset and Ground-Truth-Annotation}
This research uses a manually curated dataset of tenancy schedule PDF documents from five major real estate firms (4-6 tenancy schedules per format) to propose a pipeline for schema matching with heterogeneous table layouts. Each firm uses a distinct table layout, capturing the diversity of document formats in practice. The dataset has been fully anonymized while preserving data structure, syntax and formatting. Sensitive business data, such as tenant names and rent values have been changed to generic values. The five formats used here were: JLL, Savills, CBRE, Park15 and EDIF.

\indent Ground-truth mapping \ref{tab:fd-mapping-example} was created by manually aligning the tenancy schedules with the target schema (created with the help of the CBRE Data Requirements Documents) and grouping 20 semantically similar column sets manually. These serve as a benchmark for evaluating the accuracy of schema matching and training individual metric weights (instance- vs schema-based).

\subsection{Model Performance}

We evaluated the alignment accuracy of the template-based hybrid matcher using precision, recall, and F1-score, benchmarked against manually curated ground-truth mappings. In addition, schema usability was assessed based on schema compactness. The F1-score of the Full Disjunction (FD) baseline is not directly compared in terms of alignment accuracy, but is included to ensure that alignment quality is sufficient for a valid schema usability comparison.

\indent We also compared different configurations of the matcher, including setups with and without per-attribute optimization. For each configuration, precision, recall, and F1-score were reported to quantify matching accuracy. Precision reflects the proportion of correct predicted matches, while recall indicates how many true column-to-target mappings are included in the standardized schedule. The F1-score, representing the harmonic mean between precision and recall, served as the main metric for optimizing parameters and weights.

\indent To assess schema usability from a business perspective, we compared two configurations of the FD baseline with those of the template-based matcher. Usability was measured using two indicators: the total number of columns in the standardized schema (schema compactness) and the overall null percentage. A compact schema with fewer nulls is considered more interpretable and usable for downstream applications.

\indent Table~\ref{tab:results} summarizes the performance of all four matcher configurations compared to the FD baseline. The baseline hybrid configuration achieved an F1-score of 0.778, while applying global parameter optimization increased this to 0.800. The highest performance was achieved with per-attribute weight optimization, reaching an F1-score of 0.881. This suggests that fine-tuning metric weights at the attribute level provides a significant improvement in alignment accuracy.

\indent Although we do not evaluate the alignment accuracy of the FD baseline directly, we use its F1-score to verify whether usability metrics can be meaningfully compared. The FD baseline reached an F1-score of 0.712 and a null percentage of 75.9\%, indicating more schema bloat compared to all hybrid configurations. This supports the hypothesis that the FD approach results in reduced schema usability.

\subsection{Attribute-Level Analysis} 
To further support the analysis of hyperparameter optimization, we examined the learned relative weights for schema- and instance-based metrics across the three attribute types of STRING, DECIMAL, and DATE. Table \ref{tab:metric_gap} shows the average weight of schema- and instance-based metrics after per-attribute weight optimization using $\alpha=0.5,\theta=0.5$. 
\begin{table}[ht]
\centering
\caption{Average metric weight by data type}
\label{tab:metric_gap}
\begin{tabular}{lccc}
\toprule
\textbf{Type} & \textbf{Schema Avg} & \textbf{Instance Avg} & \textbf{Gap} \\
\midrule
STRING & 0.1500 & 0.2579 & 0.108 \\
DECIMAL & 0.1143 & 0.292 & 0.178 \\
DATE & 0.1000 & 0.356 & 0.256 \\
\bottomrule
\end{tabular}
\caption*{\textit{Weights are normalized to account for amount of metrics}}
\end{table}
The results indicate that instance-based metrics received constantly higher weights than schema-based metrics across all attribute types. The largest gain is observed for DATE attributes, where instance-based metrics outweighed schema metrics by a margin of 0.256. DECIMAL attributes also substantially benefit from instance-based features. This is likely because numerical distributions such as rent and square meters can have ambiguous header names where numerical signals are a key discriminator. This supports the idea for per-attribute weight tuning as it allows the hybrid matcher to adapt to different attribute characteristics and better use the informative signals. STRING attributes showed the smallest gap, likely due to due absence of numerical patterns to discriminate mappings. To evaluate the statistical difference between schema- and instance-based metrics, a Wilcoxon signed-rank test was performed on the metric weight configuration from the grid size 4 tuning set. The test yielded a test statistic of 3.0 with a p-value of 0.00049, this indicates a statistically significant preference for instance-based metrics.

\subsection{Error Analysis}
Despite the overall matching accuracy there were some recurring misalignments. One frequent example is the confusion between semantically similar attributes such as ``Office'', ``Office Area'', and ``Office Rent''. Without per-attribute tuned weights the matcher may incorrectly prioritize schema-based similarity (string synonym overlap) and map ``Office'' to ``Office Rent'' rather than the correct target of ``Office Area''. Per-attribute weights would reveal that the columns denotes ''Office Area'', as it fits better into the numeric distribution signals of this target attribute. This issue illustrates the risk of relying only on generic string-based metrics.

\indent These kinds of errors come to light even more in the Hungarian algorithm step. Here the algorithm seeks a globally optimal one-to-one mapping. However, even a small scoring imbalance (the difference between office\_rent and office\_area in the score is minimal) can lead to systematic mismatches when competing attributes have overlapping names. 

\indent In addition to matching errors, we observed that the overall null percentage can be a misleading statistic as there are attributes that are rarely present in any source tenancy schedules. These contribute to null values across all configurations tested. Future work could isolate new nulls introduced instead of overall null percentage.

\section{Discussion}
\label{sec:discussion}
\subsection{Interpretation of Findings}
This study set out to evaluate whether a novel template-based hybrid schema matcher could outperform an FD baseline matcher in terms of both usability and matching accuracy when aligning tenancy schedules. The results strongly support the effectiveness of the hybrid approach in terms of schema compactness and matching accuracy. Notably, the importance of per-attribute optimization is demonstrated in the performance gains across the different configurations.

\indent The core aim of this research was to introduce a business-first perspective to schema alignment which prioritized schema compactness for downstream usability. Tenancy schedules contain crucial information for real estate firms like CBRE, but their heterogeneous formats have hindered effective data mining. Automated schema matching techniques risk generating overly complex bloated schemas due to overfitting \cite{paulraj_ext-nosql_2024}. We hypothesized that a template-based approach with a target schema could result in a standardized tenancy schedule that is significantly more usable in practice.

\subsection{Theoretical Insights}
This research contributes to the general understanding of schema matching for documents of similar data content with varying layouts. We validated a novel template-based approach and a fully unsupervised automated integration method. Unlike layout-specific templates, which require manual adaptation for each layout format, the proposed hybrid matcher generalizes across layouts.

\indent The study also provides support for the idea that schema- and instance-based metrics should not be treated uniformly. We show that different data types benefit from different metrics in terms of matching accuracy. This is a nuance that general-purpose schema matchers overlook. The Hungarian algorithm offers a global optimal for one-to-one assignment. However, this is also a hard constraint that might not reflect real-world scenarios where tenancy schedule data could map to multiple target attributes or vice versa. This is a limitation which suggests a that more flexible matching algorithm could improve performance.

\subsection{Practical Implication}
Tenancy schedule data represent one of the most valuable forms of data for real estate firms like CBRE. It directly tracks leasing, asset management, and valuation data. However, while data from CBRE's internal format has been structured and utilized, tenancy schedules acquired from other firms were not analyzed and not stored as structured data in their data warehouse for downstream applications. This research addresses that gap by offering a scalable and reusable solution for organizations dealing with heterogeneous document layouts. As mentioned, real estate firms in particular can benefit from a single target schema which can be extended. This removes the need to define and maintain layout-specific templates or integration logic. 

\indent By enabling accurate schema alignment without manual rule engineering, the proposed hybrid matcher improves extendability and maintainability. It significantly reduced null density and schema size when compared to the automated baseline. For instance, the best hybrid configuration achieved an F1-score of 0.881 with a null percentage of 45.7\%, whereas ALITE, despite reaching an acceptable F1-score of 0.712 without a target schema, had an overall null percentage of 75.6\% This is a critical part when considering a more usable schema for downstream applications.

\subsection{Limitations and Future Work}
Despite the promising results presented, several limitations of this study have to be acknowledged. First, the pipeline used assumes clean, pre-extracted data. We do not handle noisy or OCR-derived data which is often encountered in real-world ingestion pipelines. Immediate applicability is limited due to this for more automated end-to-end workflows. Second, the evaluation was conducted using tenancy schedules formats from five distinct real estate firms which show representative layouts. However, it is a relatively small number. The results are therefore less generalizable. It does show the effectiveness of manually labeling, a limited set of ground-truth mappings and the resulting performance increase.

\indent Another key limitation is the assumption of one-to-one column-to-attribute in our current setup. This excludes more complex scenarios which involve composite fields which are common in practice. An example is ``Office Area'' and ``Archive Area'' merged into ``Leased Area''. Because the optimization grid space scales exponentially with the number of metrics, attributes, and grid size runtime becomes a concern. To achieve usable computing times this pipeline requires precomputing metrics and caching $.csv$ files. Currently, optimization using 16 CPU cores for 17 attributes with 7 metrics and a grid size of 5 takes 49 minutes. This demonstrates reasonable computing times are possible in the current configuration. However, as future work expands the metric set or increases the dataset and target schema, the exhaustive grid search can become prohibitive. Such cases require more efficient search strategies such as Bayesian optimization.

\indent Future research could address these limitations in several ways. First, extending this matcher to use methods for handling composite and many-to-one mappings would significantly increase the practical coverage. Second, extending the pipeline to an end-to-end pipeline with table extraction could simulate more realistic ingestion pipelines. Lastly, evaluating this proposed pipeline across other domains with heterogeneous data formats would help validate its generalizability  beyond only real estate data.

\section{Conclusion}
\label{sec:conclusion}

This research proposed a hybrid template-based schema matcher and compared it to an FD baseline (ALITE) for integrating multi-layout tenancy schedules into a target schema. This addresses a gap in existing literature where prior systems either rely on manually defined (often rule-based) templates per layout or an automated integration which produces bloated schemas. The proposed hybrid matcher, using both schema- and instance-based metrics, was evaluated with a globally optimal column-to-attribute assignment against a manually created ground-truth mapping.

\indent In addressing the evaluation of alignment accuracy and hyperparameter optimization, the proposed hybrid matcher demonstrated a high alignment accuracy to the target schema (derived from the DRD schema). The baseline configuration, using default parameters, achieved an F1-score of 0.778. After global optimization of parameters $\alpha = 0.75$, $\theta = 0.51$, performance increased a little to 0.800. The best results were obtained through per-attribute weight optimization which yielded a high F1-score of 0.881. Combining parameter optimization and metric-weight optimization proved insignificant. The overall results confirm the importance of adapting metric contributions per-attribute, Numeric fields such as ''floor-level'' benefited more from instance-based metrics while text fields such as ''tenant-name'' benefited relatively more from schema-based metrics.

\indent
This analysis focused on schema compactness and the importance of schema usability for business users. Compared to the FD baseline, which merges with an outer-join using column embedding, the hybrid matcher resulted in a compact schema (17 columns vs. 88 columns) and a lower overall null percentage in the standardized schedule ($45.7\%$ vs. $75.6\%$). These results demonstrate a significant improvement in business usability downstream by reducing overall null fields and producing a more compact schema.

\indent However, these results have several limitations. First, this study assumes clean table extraction and one-to-one column-to-attribute mapping which may not hold for more complex table structures. Second, the evaluation is limited to five different layout structures from real estate firms. Further testing (for different domains) can help improve generalization and prove the feasibility across other document types. Finally, the system depends on a well-defined target schema with clear knowledge of the required schema downstream.

\indent Despite these constraints, this research contributes a scalable template-based alternative to manual mapping and fully automated table integration. It uses a template-based approach that accommodates layout variability without layout-specific templates. Omitting layout-specific templates introduces a significant gain in re-usability and scalability in business-centric schema alignment.

\indent Future research could extend this approach to support composite and complex tables, and evaluate its generalizability beyond real estate tenancy schedules. Applying this approach to other domains such as invoices or inspection forms would help ground the broader applicability.

\bibliographystyle{ACM-Reference-Format}

\newpage
\onecolumn

\appendix
\begin{appendices}

\section{Dataset and Ground Truth Mapping}
\label{sec:apx:app_mapping}

\subsection{Ground Truth Mappings}
\begin{table}[H]
\centering
\caption{Sample of ground truth mappings used in hybrid matcher evaluation}
\label{tab:groundtruth-example}
\begin{tabular}{|l|l|l|}
\hline
\textbf{Format} & \textbf{Source Column} & \textbf{Target Attribute} \\
\hline
JLL & Office/Business space & office\_area \\
JLL & Tenant & tenant\_name \\
SAVILLS & Office space sq m & office\_area \\
SAVILLS & Total annual rent & passing\_rent\_pa \\
EDIF & Contracted Annual Rent (EUR pa) & passing\_rent\_pa \\
\hline
\end{tabular}
\end{table}

\begin{table}[H]
\centering
\caption{Sample of ground truth mappings used in FD evaluation}
\label{tab:fd-mapping-example}
\begin{tabular}{|l|l|l|l|l|l|}
\hline
\textbf{FD Set Name} & \textbf{JLL} & \textbf{Savills} & \textbf{CBRE} & \textbf{EDIF} & \textbf{PARK15} \\
\hline
Tenant Name & Tenant & Tenant & Tenant & Tenant Name & Tenant \\
Floor No & Floor & Floor & Floor(s) & Floor & NA \\
Total Leased Area (SQM) & Total & Total Area sq m & NA & NLA (Sqm) & Contractual size \\
Total Leased Area (eur) & Total.1 & Total Annual rent & Total annual rent (excl. VAT) & Contracted Annual rent & Total gross annual rent \\
Index date & first index & Next index date & Next index & NA & Next index date \\
VAT Liability (boolean) & Yes/No & VAT (liable y/n) & VAT & NA & VAT Liable \\
Parking spaces (units) & Parking & NA & PP & Parking Spaces & NA \\
\hline
\end{tabular}
\caption*{\textit{NA indicates attribute is not present in format}}
\end{table}
\noindent For full mapping set, see GitHub repository: \url{https://github.com/TUilkema}

\begin{table}[H]
    \small
    \centering
    \caption{JLL tenancy schedule format}
    \begin{tabular}{|c|c|}
    \hline
    \textbf{Column Name} & \textbf{Example Value} \\
    \hline
    Tenant & NeuroLogic 21 B.V. \\
    Floor & GF \\
    Total & 5.294 \\
    Office/Business space & 5.234 \\
    Archive & 60 \\
    Restaurant & - \\
    pp & 50 \\
    Total.1 & € 1,177,924 \\
    Office+Restaurant & € 1,100,920 \\
    Office & € 1,023,230 \\
    Archive & € 47,690 \\
    unnamed.1  & € 193 \\
    Restaurant* & € 77,690 \\
    unnamed.2 & € 24.20 \\
    Parking & € 23.000 \\
    unnamed.3 & € 460 \\
    Commencement & 6-3-2013 \\
    Expiry & 30-6-2031 \\
    Break & 30-6-2031 \\
    Tenant option(s) & 99*5 \\
    Notice Period & 12 months \\
    Notice given & No \\
    WAULT & 9.5 \\
    WAULB & 9.5 \\
    first index & 1-1-2024 \\
    Yes/No & Yes \\
    \% comp & - \\
    Guarantee & n.a. \\
    Comments & 1-5 \\
    \hline
    \end{tabular}
    \label{tab:jll}
\end{table}

\begin{table}[H]
    \small
    \centering
    \caption{Savills tenancy schedule format}
    \begin{tabular}{|c|c|}
    \hline
    \textbf{Column Name} & \textbf{Example Value} \\
    \hline
    Property code & WEB001 \\
    Property / Address & Prinsengracht 110-112 te Amsterdam \\
    Floor & -1,2,3 and 4 \\
    Unit number Savills & 0020 \\
    Tenant & Stonebridge Partners B.V. \\
    Leased space & Office \\
    Office space sq m & 620 \\
    Storage space sq m & 40 \\
    Total Area sq m & 660 \\
    Total rent office space/y & 156178,19 \\
    Total annual rent & 156178,19 \\
    Payment period (m/q) & Q \\
    VAT liable (y/n) & Y \\
    VAT comp (€) &  \\
    Start date lease & 1-6-2012 \\
    Notice period & 12 \\
    Break option date & 1-6-2022 \\
    Notice period Break option &  \\
    Expiry date & 31-5-2022 \\
    Next index date & 1-6-2025 \\
    Option period & nx5 \\
    Type of Security & Bank Guarantee \\
    Security Amount & 0,00 \\
    CPI Indices (2000=100, 2006=100, 2015=100) & 2015=100 \\
    Comments &  \\
    \hline
    \end{tabular}
    \label{tab:savills}
\end{table}

\begin{table}[H]
    \small
    \centering
    \caption{CBRE tenancy schedule format}
    \begin{tabular}{|l|l|}
    \hline
    \textbf{Column Name} & \textbf{Example Value} \\
    \hline
    Tenant & Nuvion Consulting BV \\
    Floor(s) & 4th \\
    Office (sq m) & 242 \\
    Archive (sq m) &  \\
    PP & 2 \\
    Rent office (€/sqm) & 340 \\
    Rent Archive (€/sqm) &  \\
    Rent PP (€/PP) & 2,430 \\
    Annual rent (excl. VAT) & 82,284 \\
    VAT compensation &  \\
    Total annual rent (excl. VAT) & 82,284 \\
    VAT & N \\
    Start date & 01-01-08 \\
    Next index & 01-12-12 \\
    Termination date & 30-11-15 \\
    Remaining lease term & 7.9 \\
    Notice period & 12 months \\
    Options/Extensions & N * 5 years \\
    \hline
    \end{tabular}
    \label{tab:my_label}
\end{table}

\begin{table}[H]
    \small
    \centering
    \caption{EDIF tenancy schedule format}
    \begin{tabular}{|l|l|}
    \hline
    \textbf{Column Name} & \textbf{Example Value} \\
    \hline
    EDIF ID & 0192 \\
    Property ID No. & P006 \\
    Property Name & Apeldoorn \\
    Country & Netherlands \\
    Demise ID No. & D001 \\
    Floor & 0 \\
    Tenant ID No. & T0054 \\
    Tenant Name & Enovix Utilities B.V. \\
    Use & Office \\
    NLA (Sqm) & 1.409 \\
    Parking Spaces & - \\
    Lease Start Date & 1-jan-2016 \\
    Break Date & - \\
    Expiry Date & 31-dec-2024 \\
    Earliest Expiry Date & 31-dec-2024 \\
    Contracted Rent at Reporting Date (€ psqm pm) & 6,29 \\
    Contracted Rent at Reporting Date (€ per unit pm) &  \\
    Contracted Annual Rent (€ pa) & 106.351 \\
    TOTAL & 109.442 \\
    14,50\% & 125.858 \\
    \hline
    \end{tabular}
    \label{tab:edif}
\end{table}

\begin{table}[H]
    \small
    \centering
    \caption{PARK15 tenancy schedule format}
    \begin{tabular}{|l|l|}
    \hline
    \textbf{Column Name} & \textbf{Example Value} \\
    \hline
    ID & 1 \\
    Address & Parkweg 2 \\
    Tenant & IKEA B.V. \\
    Brand & IKEA \\
    Contractual size (sqm) & 2550 \\
    LFA/GFA & - \\
    Total size (sqm LFA NEN2580) & 674 \\
    Start date & 01-01-2012 \\
    Expiry date & 31-12-2024 \\
    Break date & - \\
    WALL (to break) & 5.5 \\
    WALL (to expiry) & 5.5 \\
    Option periods & 1*5 \\
    Extension periods & n*5 \\
    Notice periods (months) & 12 \\
    Total gross annual rent & € 385,416 \\
    Next index date & 01-01-2025 \\
    VAT liable & Yes \\
    Type & Deposit \\
    Amount & € 40,000 \\
    Terminated lease & No \\
    Comments & - \\
    \hline
    \end{tabular}
    \label{tab:park15}
\end{table}

\section{Metrics}
\label{app_metrics}
\subsection{Schema-Based Metrics}
\textbf{Levenshtein Similarity:}
Measure the normalized edit distance between the cleaned source column header and the target attribute's synonyms.

\textbf{Formula:}
\[\text{Sim}_{\text{lev}}(s_1, s_2) = \max \left(1 - \frac{\text{EditDist}(s_1, s_2)}{\max(|s_1|, |s_2|)}\right)\]

\textbf{Jaccard Token Overlap:}
Measure set-based token similarity between column header and each synonym of target attribute.

\textbf{Formula:}
\[\text{Sim}_{\text{jaccard}}(T_1, T_2) = \frac{|T_1 \cap T_2|}{|T_1 \cup T_2|}\]\[\text{Schema}_{Jaccard}(h, a) = \max_{s_2 \in \{\text{name, synonyms}\}}\text{Sim}_{\text{jaccard}}(T_1, T_2)\]

\textbf{Synonym Matching:}
Use fuzzy token-set ratio to compare column header with all known synonyms of target attribute.

\textbf{Formula:}
\[\text{Sim}_{Syn}(h, a) = \max_{s \in \{\text{name, synonyms}\}} \frac{\text{FZS}(h, s)}{100}\]
\textit{*FZS = Fuzzy Token Set}

\subsection{Instance-Based Metrics}
\textbf{Numeric Data Type:}
Estimates how likely the column is to contain numerical data, based on content of its cell.
This metric checks (1) how many cells contains a digit and (2) how many cells start with a digit. Combines this in a weighted score.

\textbf{Formula:}
\[N(C) = 0.7 \cdot \text{HasDigit}(C) + 0.3 \cdot \text{StartsWDigit}(C)\]

\textbf{Date Type:}
Calculates the proportion of cells that match into a date pattern.

\textbf{Formula:}
\[\text{Sim}_{Date} = \frac{1}{N} \sum_{i=1}^N \mathbb{1}_{\text{is\_date}}(d_i)\]

\textbf{Numerical Range Comparison:}
Evaluate how close the mean of source column is to mean of target attribute and how many values lie within IQR.

\textbf{Formula:}
\[\text{Sim\_Range} = 0.5 \cdot \max\left(0, 1 - \frac{|\mu_s - \mu_t|}{\text{IQR}}\right) + 0.5 \cdot \frac{1}{N} \sum_{i=1}^N \mathbb{1}_{[q_1, q_3]}(x_i)\]

\textbf{KS-Test (Kolmogorov–Smirnov test):}
Compares the distribution of the source column to a normal distribution generated using the target's mean and IQR.

\textbf{Formula:}
\[D = \sup_x |F_s(x) - F_t(x)| \quad \text{where } F_s, F_t \text{ are empirical CDFs}\]
\[
\text{KS\_Score} = 1 - D
\]
\textit{CDF = Cumulative Distribution Function}

\end{appendices}

\end{document}